\documentclass[english,reprint,amsmaths,amssymb,prl]{revtex4-1}
\usepackage[T1]{fontenc}
\usepackage[latin9]{inputenc}
\usepackage{color}
\usepackage{refstyle}
\usepackage{amstext}
\usepackage{graphicx}
\usepackage[unicode=true,pdfusetitle,
 bookmarks=false,
 breaklinks=false,pdfborder={0 0 0},pdfborderstyle={},backref=false,colorlinks=true]
 {hyperref}
\hypersetup{
 pdfborderstyle={}}

\makeatletter

\AtBeginDocument{\providecommand\tabref[1]{\ref{tab:#1}}}
\AtBeginDocument{\providecommand\figref[1]{\ref{fig:#1}}}
\providecommand{\tabularnewline}{\\}
\RS@ifundefined{subsecref}
  {\newref{subsec}{name = \RSsectxt}}
  {}
\RS@ifundefined{thmref}
  {\def\RSthmtxt{theorem~}\newref{thm}{name = \RSthmtxt}}
  {}
\RS@ifundefined{lemref}
  {\def\RSlemtxt{lemma~}\newref{lem}{name = \RSlemtxt}}
  {}

\renewcommand{\tabref}{\Tabref}
\renewcommand{\figref}{\Figref}
\usepackage[shortcuts]{extdash}

\makeatother

\begin{document}

\title{Solving the electronic structure problem for over 100,000 atoms in
real-space}

\author{Mehmet Dogan,$^{1}$ Kai-Hsin Liou$^{2}$ and James R. Chelikowsky$^{1,2,3}$}

\affiliation{$^{1}$Center for Computational Materials, Oden Institute for Computational Engineering and Sciences, University of Texas at
Austin, Texas 78712, USA $\linebreak$ $^{2}$McKetta Department of
Chemical Engineering, University of Texas at Austin, Texas 78712,
USA $\linebreak$ $^{3}$Department of Physics, University of Texas
at Austin, Texas 78712, USA}
\begin{abstract}
Using a real-space high order finite-difference approach, we investigate the
electronic structure of large spherical silicon nanoclusters. Within Kohn\textendash Sham density functional theory and using pseudopotentials, we report the self-consistent field convergence of a system with over 100,000 atoms: a Si$_{107,641}$H$_{9,084}$ nanocluster with a diameter of
16 nm. Our approach uses Chebyshev filtered subspace iteration to speed-up the convergence of the eigenspace, and blockwise Hilbert space filling curves to speed-up sparse matrix\textendash vector multiplications, all of which is implemented in the PARSEC code. For the largest system, we utilized 2048 nodes
(114,688 processors) on the Frontera machine in the Texas Advanced
Computing Center. Our quantitative analysis of the electronic structure
shows how it gradually approaches its bulk counterpart as a function
of the nanocluster size. The band gap is enlarged due to quantum confinement
in nanoclusters, but decreases as the system size increases, as expected.
Our work serves as a proof-of-concept for the capacity of the real-space
approach in efficiently parallelizing very large calculations using high performance computer platforms, which can straightforwardly be replicated in other systems with more than $10^{5}$ atoms.
\end{abstract}
\maketitle
Calculating the electronic structure of materials has been a primary
aim for the theory of condensed matter physics, as well as the starting
point for calculating materials properties. The development of density
functional theory (DFT) \citep{hohenberg1964inhomogeneous,kohn1965selfconsistent}
and pseudopotential methods \citep{chelikowsky1992abinitio} made
it possible to treat the problem as an effective one-electron problem,
rather than the much more difficult interacting many-electron problem.

Over the past few decades, rapid improvements in computational capabilities
allowed researchers to solve the electronic structure problem for
larger and larger systems using various software packages. Currently,
the most common available packages use a plane wave basis in which
to expand the Kohn\textendash Sham wave functions. Despite the practical
success of this approach, a few disadvantages exist: (1) the requirement
of extensive global communications hindering massive parallelization;
(2) the inability to directly calculate aperiodic structures such
as nanoclusters and instead having to construct large supercells;
and (3) the inability to calculate charges systems without a compensating
background charge that might alter their properties.

An alternative approach utilizes a real-space grid to circumvent these
shortcomings \citep{chelikowsky1994finitedifferencepseudopotential,stathopoulos2000parallel,chelikowsky2003usingreal,alemany2004realspace}.
In our approach, which has been implemented in the software package
PARSEC (``pseudopotential algorithm for real-space electronic
structure calculations''), the Kohn\textendash Sham equations are
solved \emph{via} high order finite difference in real-space \citep{kronik2006parsectextendash,gavini2022roadmap}.
The reduced need for global communication makes parallelization easier,
and periodic simulation cells are not required (overcoming the latter
two aforementioned issues), though they are also implemented. Various other
implementations in real-space have also been developed, \emph{e.g.}
multigrids \citep{bernholc2008recentdevelopments,cohen2013locally,zhang2019largescale},
multiwavelets \citep{jensen2017theelephant}, finite-element \citep{pask2001finiteelement,kanungo2019realtime})
as well as other finite-difference implementations \citep{iwata2010amassivelyparallel,andrade2015realspace,mi2016atlasa,michaud-rioux2016rescua,ghosh2017sparcaccurate,ghosh2017sparcaccurate2}.
Achieving convergence within the finite-difference formalism is straightforward if the  Coulomb singularity of the all electron potential is removed using pseudopotentials \citep{chelikowsky2000thepseudopotentialdensity}. Additionally, the
Hamiltonian matrices obtained on the real-space grid are large, but very
sparse, enabling an efficient diagonalization \citep{saad2010numerical}.

Achieving self-consistency in large systems often involves handling a bottleneck associated with the solution 
 of the eigenvalue problem.  
If there were a way to approximate the solution of the eigenvalue
problem for the first few self-consistent field (SCF) iterations without
attempting full accuracy for individual eigenstates, that would result in
a significant speed-up. This idea has been realized by the Chebyshev-filtered
subspace iteration (CheFSI).  The focus of CheFSI is   on improving the subspace
(and as a result, the charge density) and the potentials simultaneously
over successive SCF iterations \citep{zhou2006parallel,chelikowsky2007algorithms,zhou2014chebyshevfiltered,liou2020scalable}.
Rather than searching for individual eigenstates, the CheFSI method
searches for an invariant subspace, which gradually approaches the SCF
eigen-subspace. The CheFSI method provides a significant speed-up
(up to tenfold) compared to standard diagonalization techniques \citep{zhou2006parallel,liou2020scalable},
which has led to its widespread use in real-space DFT packages \citep{michaud-rioux2016rescua,ghosh2017sparcaccurate,ghosh2017sparcaccurate2,kanungo2017largescale,banerjee2018twolevel}.
The details of our current implementation of CheFSI in PARSEC can
be found elsewhere~\citep{liou2020scalable}.

\begin{table*}
\begin{centering}
\begin{tabular}{|c|c|c|c|c|c|c|c|c|}
\hline 
 & $h$ (bohr) & Machine & \#nodes & Diameter (nm) & \#grid points & \#states & \#SCF steps & walltime (h)\tabularnewline
\hline 
\hline 
Si$_{1,947}$H$_{604}$ & 0.7 & Cori & 2 & 4.2 & 1,575,600 & 4,800 & 17 & 2.5\tabularnewline
\hline 
Si$_{4,001}$H$_{1,012}$ & 0.7 & Cori & 8 & 5.3 & 2,707,504 & 9,216 & 19 & 4.0\tabularnewline
\hline 
Si$_{10,869}$H$_{1,924}$ & 0.7 & Cori & 64 & 7.5 & 6,377,184 & 24,576 & 17 & 8.4\tabularnewline
\hline 
Si$_{23,049}$H$_{3,220}$ & 0.7 & Cori & 256 & 9.6 & 15,180,904 & 61,440 & 18 & 27.9\tabularnewline
\hline 
Si$_{51,071}$H$_{5,484}$ & 0.9 & Frontera & 512 & 12.5 & 15,522,368 & 114,688 & 15 & 28.6\tabularnewline
\hline 
Si$_{107,641}$H$_{9,084}$ & 0.9 & Frontera & 2048 & 16.0 & 31,901,640 & 245,760 & 14 & 46.8\tabularnewline
\hline 
\end{tabular}
\par\end{centering}
\caption{\label{tab:Summary}The silicon NCs studied in this work. For each
system, the corresponding grid spacing ($h$), the machine on which
the calculation was run, the number of computing nodes, the diameter
of the NC, the number of grid points, the number of computed states,
the number of SCF steps to reach convergence, and the walltime for
the calculation are tabulated.}
\end{table*}

The efficiency of the CheFSI method relies on sparse matrix\textendash vector
multiplications arising from the Hamiltonian and the wave functions.
In order to parallelize these multiplications, one needs to partition
the calculational domain and distribute among processors, which has
the danger of being very inefficient when thousands of processors
are involved. We have recently investigated the most efficient ways
of domain partitioning based on space filling curves (SFCs)~\citep{liou2021spacefilling}.
SFCs are continuous curves that traverse a 3D domain and pass
through every grid point once. This presents a straightforward way
to access data on each point, while the self-similarity of SFCs preserve
the locality of the grid points on the curve \citep{lawder2000usingspacefilling,moon2001analysis,yzelman2014highlevel,xu2014optimality}.
We have found that using blockwise Hilbert SFCs can provide an over
six-fold speed-up of sparse matrix\textendash vector multiplications
compared to simple Cartesian ordering \citep{liou2021spacefilling}.

Here, we present a large Kohn\textendash Sham DFT calculation
 with SCF convergence (a silicon nanocluster made up of over
100,000 atoms) using the Chebyshev-filtered subspace iteration and
blockwise Hilbert space-filling curves implemented in the real-space
code PARSEC. We also compare this system (Si$_{107,641}$H$_{9,084}$)
to smaller nanoclusters to investigate the evolution of the electronic
structure as a function of the system size. This achievement
demonstrates the robust scalability and parallelizability of the real-space
method.

The calculations presented here were run on two supercomputers: the
National Energy Research Scientific Computing Center's (NERSC) Cori
machine, and the Texas Advanced Computing Center's (TACC) Frontera machine.
On Cori, we used Knights Landing (KNL) nodes, each of which is equipped
with one Intel Xeon Phi 7250 processor, which has 68 cores. On Frontera,
each node is equipped with one Intel Xeon Platinum 8280 processor,
which has 56 cores. We used norm-conserving pseudopotentials, constructed
by the Troullier\textendash Martins method in the Kleimann\textendash Bylander
form \citep{troullier1991efficient,kleinman1982efficacious}. We employed the
 local density approximation (LDA) to approximate the exchange\textendash correlation functional \citep{ceperley1980groundstate,perdew1981selfinteraction}.
We applied four iterations of CheFSI during the first SCF step, and
then one iteration for each following step. Our SCF convergence criterion
is SRE $<0.0001$ Ry, where SRE (self-consistent residual error) is
defined as the integral of the square of the difference between the
last two self-consistent potentials, weighted by electron density
and taken squared root.

We list and summarize the calculations we ran in \tabref{Summary}.
Each calculation is a spherical silicon nanocluster (NC) generated
from bulk silicon in the diamond crystal structure with the optimized
lattice constant 5.38~$\mathring{\text{A}}$, within one percent of the observed lattice constant of 5.43$~\mathring{\text{A}}$. The dangling bonds
of the surface silicons are passivated by hydrogen atoms. Among the
six NCs listed, the second, third and fourth ones were previously
reported in ref. \citep{liou2021spacefilling}, and are included here
for comparison. For the larger NCs (Si$_{51,071}$H$_{5,484}$ and
Si$_{107,641}$H$_{9,084}$), we switched from the Cori machine to
the Frontera machine, which has a higher clock rate (2,7 GHz vs. 1.4
GHz) and memory per node (192 GB vs. 96 GB). For these NCs, we also
increased the grid spacing $h$ from 0.7 bohr to 0.9 bohr, which reduced
the problem size by a factor of $\left(0.7/0.9\right)^{3}\simeq0.47$.
We have verified for the smaller NCs that the grid spacing of $0.9$
bohr leads to well-converged eigenvalues for the energy spectrum,
leading to histograms of energies hardly distinguishable from those
resulting from $h=0.7$ bohr. We note that the largest NC studied
here required the use of 2048 nodes (114,688 cores), which corresponds to a quarter of the full Frontera capacity. In comparison, the only other DFT calculation of a similar size to our knowledge, which is a single SCF iteration on a 107,292-atom Silicon nanowire (also utilizing high order finite-difference
real-space methods) was achieved using 442,384 cores on the K Computer at Riken Advanced Institute for Computational Science \citep{hasegawa2011firstprinciples}. 

\begin{figure}
\centering{}\includegraphics[width=0.9\columnwidth]{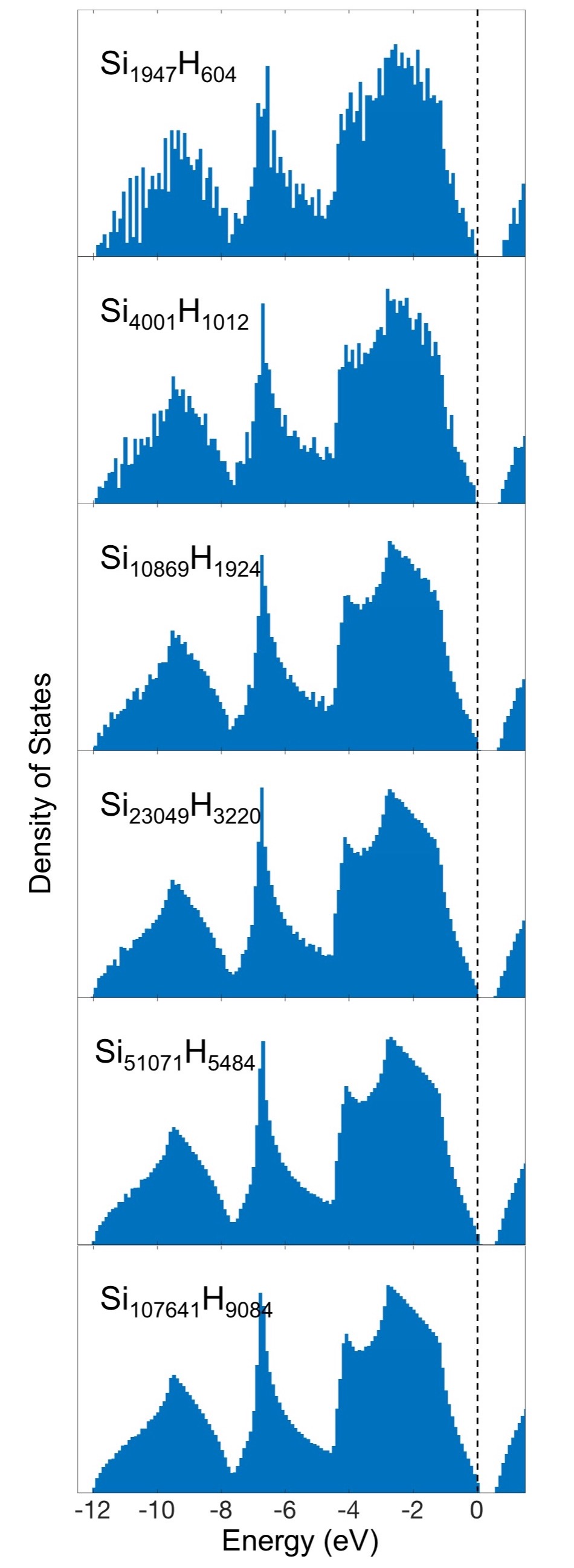}\caption{\label{fig:DOSsequence}The density of states (DOS) for all of the
six NCs investigated in this study. The DOS for each case is obtained
by plotting a histogram of the eigenstates using 0.1 eV bins. The
highest occupied state (Fermi level) is set to zero in each case.}
\end{figure}

In real-space, each state is a function of space in the whole calculation domain. Therefore, the total number of degrees of freedom in our largest calculation is $7.8\times 10^{12}$ (computed as \#grid points $\times$ \#states read off \tabref{Summary}). The fact that the Hamiltonian is sparse and we are only computing a small low-energy subset of the eigenspace (0.77\% of the total number of eigenstates) greatly helps in making this problem more tractable \citep{schofield_spectrum_2012}.

We present the density of states computed for each of the six NCs
in \figref{DOSsequence}. In order to plot the DOS for a given system,
we first set the eigenvalue for the highest occupied state (Fermi
level) to zero, then create a histogram of eigenvalues using 0.1 eV
bins, and finally normalize each plot so that they all have the same
total area below the Fermi level. We observe that as the NC gets larger,
the DOS becomes less ``noisy,'' and the familiar features of the bulk
DOS become more discernible. The dip around $-8$ eV and the van Hove
singularity around $-7$ eV become sharper for larger NCs \citep{cohen1988electronic}.
We note that these sharp features which can be understood
in the context of band theory nevertheless arise in our real-space
calculation which does not invoke Bloch's theorem.

\begin{figure}
\centering{}\includegraphics[width=0.9\columnwidth]{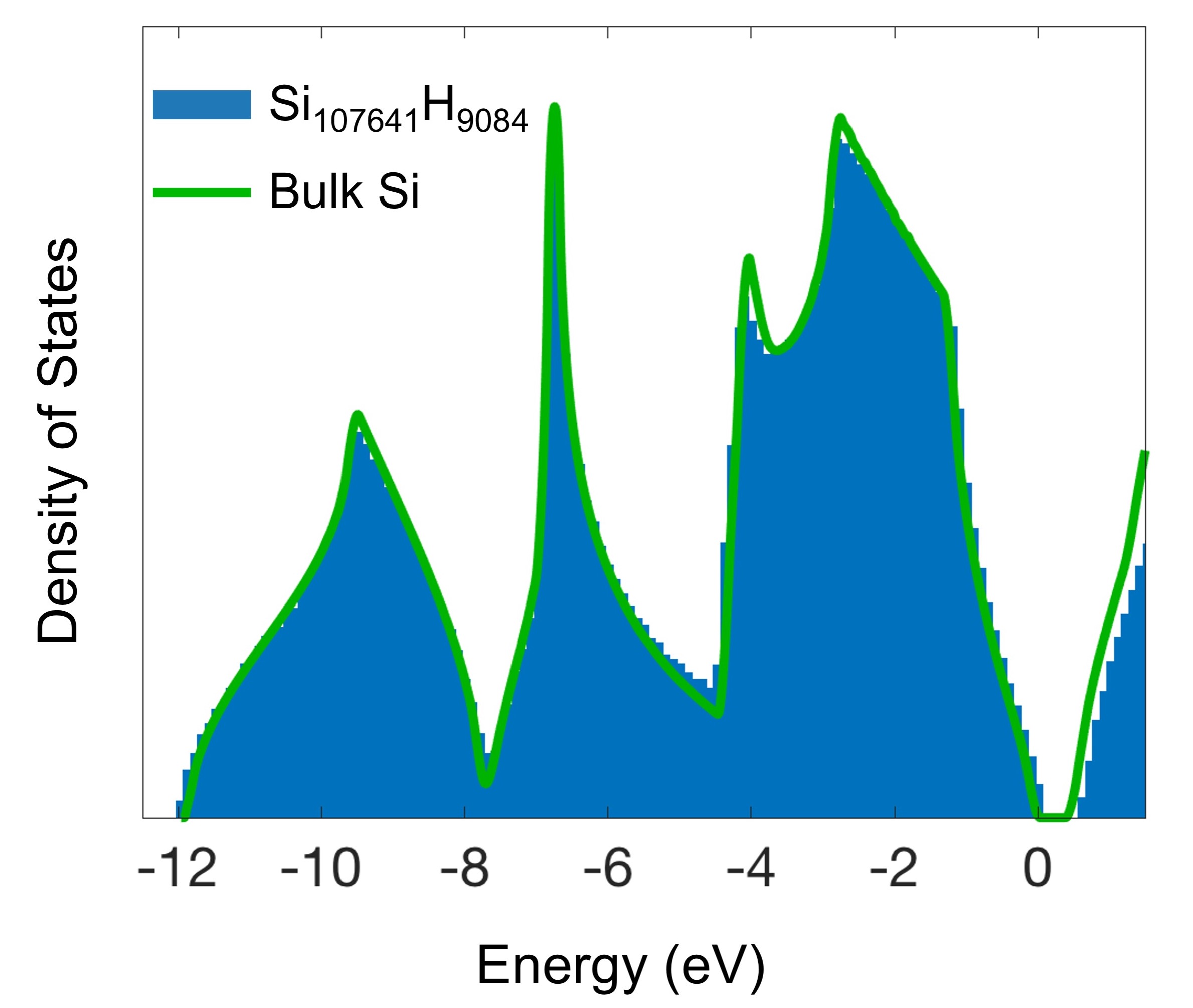}\caption{\label{fig:DOS100k}The density of states (DOS) for the Si$_{107,641}$H$_{9,084}$
NC (blue bars) plotted together with the DOS of bulk silicon (green
line), both obtained from PARSEC using similar grid spacings. The
NC DOS is obtained by plotting a histogram of the eigenstates using
0.1 eV bins, and then normalized to match the area under the valence
band part of the bulk Si curve. The highest occupied state of the NC is
set to zero, and the valence band edge of the bulk DOS is positioned
to minimize the RMS difference between the two plots.}
\end{figure}

In order to systematically study how the electronic structure of the
NCs converge to the bulk, we performed a bulk silicon calculation
in PARSEC with $h=0.86$ using the periodic boundary conditions \citep{alemany2004realspace,natan2008realspace}.
and a $36\times36\times36$ Monkhorst\textendash Pack \emph{k}-point
sampling for the 8-atom simple cubic cell \citep{monkhorst1976special}.
To compare the bulk DOS and the DOS of the Si$_{107,641}$H$_{9,084}$
NC, we first sample the bulk DOS on the same energy values that are
used to generate the DOS of the NC. We then normalize the DOS of the
NC such that the total area under the valence band part is equal to
the bulk counterpart. We then calculate the root mean square (RMS)
of the difference between the two curves, which we self-consistently
minimize by horizontally shifting the bulk DOS plot. The resulting
comparison with a minimized RMS is presented in \figref{DOS100k}.
All features of the bulk DOS are replicated in the NC DOS with high
precision. The deviation at the conduction band edge is an expected
result of quantum confinement.

\begin{figure}
\centering{}\includegraphics[width=0.9\columnwidth]{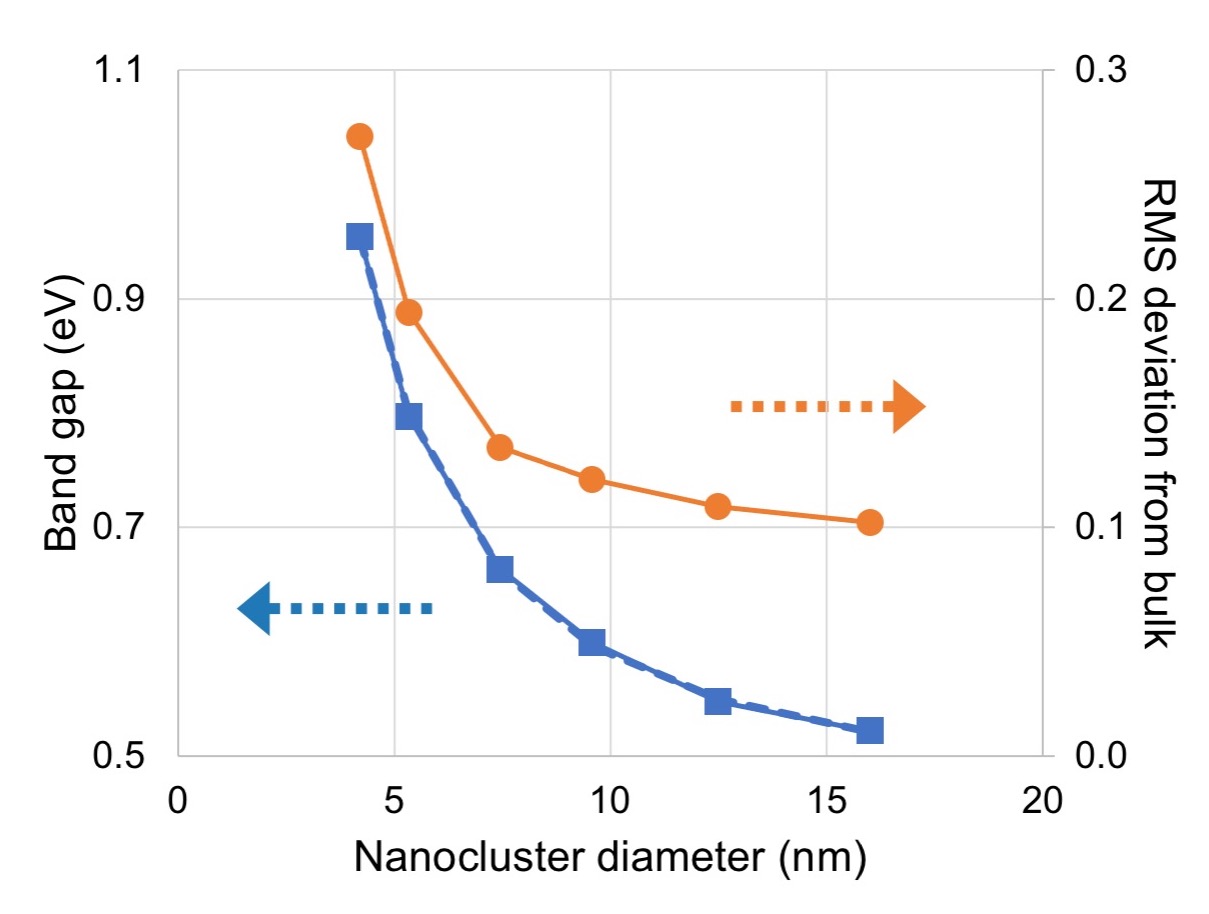}\caption{\label{fig:Plots} The evolution of the band gap and the RMS deviation
for each DOS from the bulk DOS for the six computed silicon NCs considered
in this work. The blue curve with square marks plots the band gap
(left axis), and the orange curve with round marks plots the TMS deviation
from the bulk DOS (right axis). A best fit curve with a power law
dependence and a vertical offset is also included for the band gap
curve (dashed). The fit has $R^{2}=0.9998$.}
\end{figure}

By repeating the same analysis for the smaller NCs, we have obtained
\figref{Plots}, which shows the evolution of the band gap and the
RMS deviation for each DOS from the bulk DOS. We observe that the
deviation from bulk electronic structure monotonically decreases with
the size of the NC, the decrease being sharper up to $\sim10$ nm.
The evolution of the fundamental band gap (\emph{i.e. }ground state
HOMO\textendash LUMO gap) follows a power law, in accordance with
previous studies on quantum confinement \citep{brus1984electrontextendashelectron,brus1986electronic,delerue1993theoretical,ougut1997quantum,kocevski2013transition,barbagiovanni2014quantum}.
By fitting the band gap \emph{vs.} diameter $d$ to the expected equation,
we find 
\[
E_{\text{gap}}\left(d\right)=0.46\text{ eV}+\frac{4.64}{\left(d\text{ in nm}\right)^{1.56}}\text{ eV},
\]
with a coefficient of determination of $0.99$98 (plotted with a blue
dashed line in \figref{Plots}). The 0.46 eV value also agrees with the band gap computed in our bulk Si calculation. Previous studies into smaller Si
NCs have found smaller exponents (between 1.1 and 1.39) \citep{delerue1993theoretical,ougut1997quantum,kocevski2013transition}.
The fact that our exponent is larger is in agreement with the expectation
that in the bulk limit, the effective mass theory of quantum confinement
becomes valid, and the exponent approaches 2 \citep{brus1984electrontextendashelectron,brus1986electronic,delerue1993theoretical}.

In summary, we have presented the successful SCF convergence of one of the 
largest systems to date (Si$_{107,641}$H$_{9,084}$ nanocluster
with a diameter of 16 nm) in Kohn\textendash Sham density functional
theory, achieved using a real-space finite-difference approach implemented
in the PARSEC code. Our method utilizes Chebyshev filtered subspace
iteration to speed-up the convergence of the eigenspace, and blockwise
Hilbert space filling curves to speed-up sparse matrix\textendash vector
multiplications. For this work, we executed highly-parallelized runs
with up to 2048 nodes (114,688 cores) on TACC Frontera. We have
also systematically investigated the convergence of the electronic
structure into its bulk counterpart as a function of the nanocluster
size. We have confirmed the predicted behavior of the band gap as
it is enlarged due to quantum confinement in nanoclusters. Our work
demonstrates the capabilities of the real-space approach in achieving
a very high level of parallelization in modern supercomputers, delivering
accurate electronic structure results in unprecedented system sizes.
As our supercomputers continue to improve into the so-called exascale
era, the ever-growing electronic structure community needs software
that can scale up and meed the challenge. We have shown that fully
\emph{ab initio} calculations in the $\left(\sim10\text{\ nm}\right)^{3}$
scale can now be done, which will pave the way for further applications
in the exascale era.

\section*{Acknowledgments}

MD acknowledges support from the \textquotedblleft Characteristic
Science Applications for the Leadership Class Computing Facility\textquotedblright{} project, which is supported by National Science Foundation award \#2139536. JRC and KHL acknowledge support by a subaward from the Center for Computational Study of Excited-State Phenomena in Energy Materials (C2SEPEM) at the Lawrence Berkeley National Laboratory, which is funded by the U.S. Department of Energy under contract no. DE-AC02- 05CH11231, as
part of the Computational Materials Sciences Program. Computational
resources are provided by the National Energy Research Scientific
Computing Center (NERSC) as well as the Texas Advanced Computing Center
(TACC).  JRC also acknowledges support from the Welch Foundation under grant F-2094. We thank Dr. Junjie Li from TACC for the excellent technical support.

\bibliographystyle{apsrev4-1}
\bibliography{Citations}

\end{document}